\documentclass[]{spie}  

\usepackage{amsmath,amsfonts,amssymb}
\usepackage{graphicx}
\usepackage[colorlinks=true, allcolors=blue]{hyperref}

\title{Measurement of All-sky Neutral hydrogen Absorption Spectrum (MANAS)}

\author[a,*]{Vinand Prayag}
\author[a,b,c,*]{Nivedita Mahesh}
\author[a]{Gregg Hallinan}
\author[a]{Yanfen Lin}
\author[a]{Joseph Lazio}
\author[]{Andrew Romero-Wolf}
\author[b]{Judd Bowman}
\author[a]{Mike Virgin}
\author[a]{Charlie Harnach}
\author[a]{Mark Hodges}

\affil[a]{California Institute of Technology, 1200 E California Blvd, Pasadena, USA}
\affil[b]{Arizona State University, 411 N Central Ave, Phoenix, USA}
\affil[c]{University of Colorado Boulder, UCB 389, Boulder, CO 80309-0593, USA}

\authorinfo{$^{*}$These authors contributed equally to this work; please cite as Prayag \& Mahesh et al. (2026).\\Further author information: \\Vinand Prayag: E-mail: vinand@caltech.edu, Telephone: +17603586403\\  Nivedita Mahesh: E-mail: nmahesh@caltech.edu}

\begin{document} 
\maketitle

\begin{abstract}
MANAS is a single antenna, ground-based platform built to measure the sky averaged ('global') redshifted 21-cm signal from Cosmic Dawn. The cosmological signal is expected as a $\approx$ 100–200 mK absorption trough against a foreground of 3-4 orders of magnitude brighter. The current crop of 21-cm experiments are limited by systematics namely the sky, ionosphere, beam, and signal path effects. 

MANAS combines an achromatic monopole antenna, an absolutely-calibrated receiver, and in-situ beam mapping with the OVRO-LWA, and develops an end-to-end error budget that quantifies each systematic. 
\end{abstract}


\section{INTRODUCTION}
\label{sec:intro}  

The redshifted 21-cm hyperfine spin-flip from neutral hydrogen is a powerful probe of the early Universe spanning the Dark Ages, cosmic dawn, and the subsequent epoch
of reionization (EoR) \cite{pritchard_constraining_2010}. Its strength is governed by: abundance of the neutral hydrogen atoms and the coupling
between the spin temperature ($T_{s}$) of the hyperfine transition and the gas temperature. 

During cosmic dawn, Lyman-$\alpha$ photons from the first luminous sources couple $T_{s}$ to the colder gas through the Wouthuysen-Field effect\cite{wouthuysen_excitation_1952,field_excitation_1958}, driving the line into absorption against the CMB; subsequent X-ray heating can later drive it into
emission. The signal spans roughly $6 \lesssim z \lesssim 37$ (about 35--200 MHz); at lower frequencies the physics moves into the dark ages, where ionospheric absorption makes ground-based detection impractical. Experiments targeting the dark ages are proposed as space missions either on the moon or in orbit\cite{jones_dark_2015}. 

The cosmic dawn signal is given as a differential brightness temperature ($T_{b}$) which is strongly model dependent. Fiducial stellar population models give estimates of the order of -80 $\lesssim$ $T_{b}$ $\lesssim$ -107 mK. However, this estimate could be larger depending on the X-ray and Ly$\alpha$ emittivity of the first luminous sources\cite{furlanetto_cosmology_2006,pritchard_21-cm_2012}. For a credible detection to be possible, the signal will have to be measured against a foreground that is 3 to 4 orders of magnitude stronger\cite{mozdzen_limits_2016}. Two methods are available for such detection. The statistical fluctuations in the 3D brightness temperature, called the 21 cm power spectrum, can be measured. This would trace the evolution of the EoR as more neutral hydrogen gets ionized creating HII regions or bubbles throughout the epoch. Interferometric instruments such as HERA\cite{deboer_hydrogen_2017}, LOFAR\cite{harker_power_2010} and MWA\cite{yoshiura_new_2021} have attempted to make a map of the intergalactic medium by making tomographic measurements.\cite{pritchard_constraining_2010}.  The other method would be to measure the differential brightness temperature over the entire surveyed sky which is called the "global" 21-cm signal. This can be achieved using single dipole antennas exemplified by experiments such as the Experiment to Detect the Global EoR Signature (EDGES)\cite{Bowman:2007su}, Shaped Antenna measurement of the background RAdio Spectrum (SARAS)\cite{2021arXiv210401756N} and Broadband Instrument for Global HydrOgen ReioNisation Signal (BIGHORNS)\cite{2015PASA...32....4S}.

The dominant obstacle for global measurements is instrument chromaticity, which can
readily mimic the cosmic-dawn signal; the least chromatic broadband, wide-field
antennas are therefore critical for foreground removal while remaining simple to
engineer \cite{mahesh_validation_2021}. The Measurement of All-sky Neutral hydrogen Absorption Spectrum (MANAS) is a single antenna ground-based hardware platform built to detect the global 21-cm signal by developing an error budget along the way that characterizes the contribution of each systematic that might corrupt the signal integrity. 

\section{MOTIVATION FOR MANAS}
The EDGES collaboration reported the first tentative detection of the redshifted 21-cm absorption signal from cosmic dawn\cite{bowman_detection_2018}. But its anomalous depth has not been reproduced, most notably in the SARAS-3 measurement\cite{singh_detection_2022}. This leaves open whether the feature is cosmological or an uncorrected instrumental systematic. Resolving this calls for an independent measurement using a different engineering approach, in which the dominant systematics: antenna chromaticity, absolute calibration, and foreground separation, are characterized rather than modeled.

MANAS was built for this purpose. It combines the strongest elements of three current 21-cm experiments: the achromatic antenna of SARAS, the high-fidelity receiver calibration of EDGES, and in-situ beam mapping with the Owens Valley Radio Observatory Long Wavelength Array (OVRO-LWA)\cite{eastwood_radio_2018}.  Operating as an outrigger to the array brings a further advantage: OVRO-LWA continuously characterizes the ionosphere over the same sky, so these time-variable effects can be measured in situ and folded into the MANAS error budget rather than left as an unconstrained systematic.

Current sky maps produced by OVRO-LWA in the 30–80 MHz range carry an absolute flux calibration uncertainty of 10–15\%, which is based on sky models and bright astronomical sources to set the flux scale. The MANAS receiver, on the other hand is calibrated against physical blackbody loads at known thermodynamic temperatures, bypassing the need for astrophysical sources. The receiver has been used to correct existing sky maps at 45 and 150 MHz which reduced the absolute calibration uncertainty to 2–5\%\cite{monsalve_absolute_2021}. MANAS thus serves as an absolute flux reference for OVRO-LWA, tying its 30–88 MHz sky maps to a known temperature scale, reducing their flux-scale uncertainty from 10–15\% to a few percent.

\section{INSTRUMENT OVERVIEW}
\label{sec:title}

MANAS operates as an outrigger radiometer to the OVRO-LWA, occupying the array position formerly used by the LEDA experiment (LWA element 252). The antenna is designed to operate over 30–100 MHz; intense RFI in Owens Valley restricts the usable science band to 30–88 MHz, the upper edge matching that of the OVRO-LWA.
This band corresponds to the redshifted 21-cm line over $15 \lesssim z \lesssim 46$

\begin{figure}[h!]
    \centering
    \includegraphics[width=0.5\linewidth]{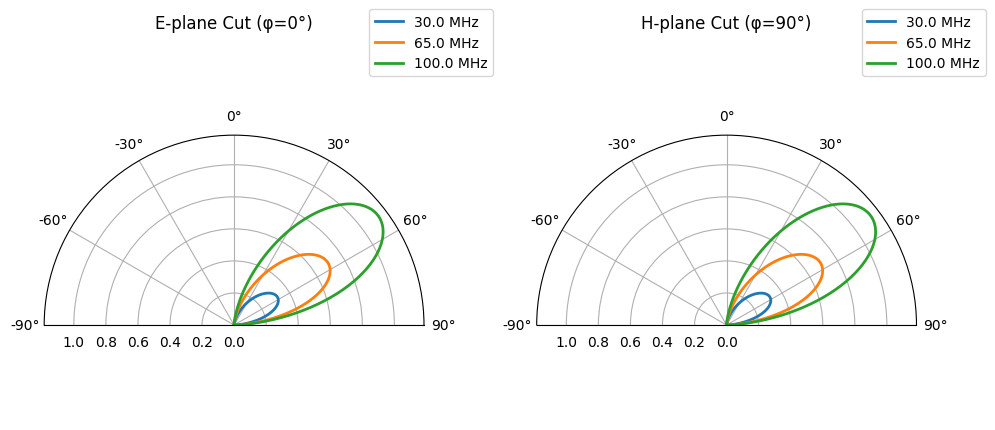}
    \includegraphics[width=0.4\linewidth]{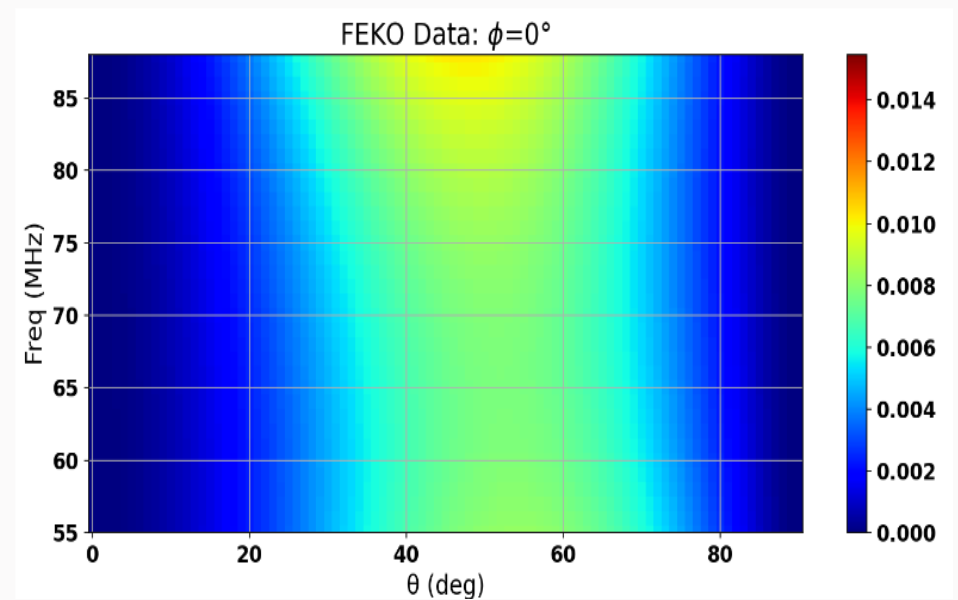}
    \caption{a.) Simulated MANAS beam (FEKO). E-plane ($\phi=0^\circ$) and H-plane ($\phi=90^\circ$) cuts at 30, 65, and 100~MHz, showing a zenith-pointing beam that broadens with frequency. b.) Simulated MANAS beam (FEKO) in the $\phi=0^\circ$ plane versus zenith angle $\theta$ and frequency. The beam varies smoothly across the band, with no resonant structure.}
    \label{fig:beamcuts}
\end{figure}

\begin{figure}[h]
    \centering
    \includegraphics[width=0.6\linewidth]{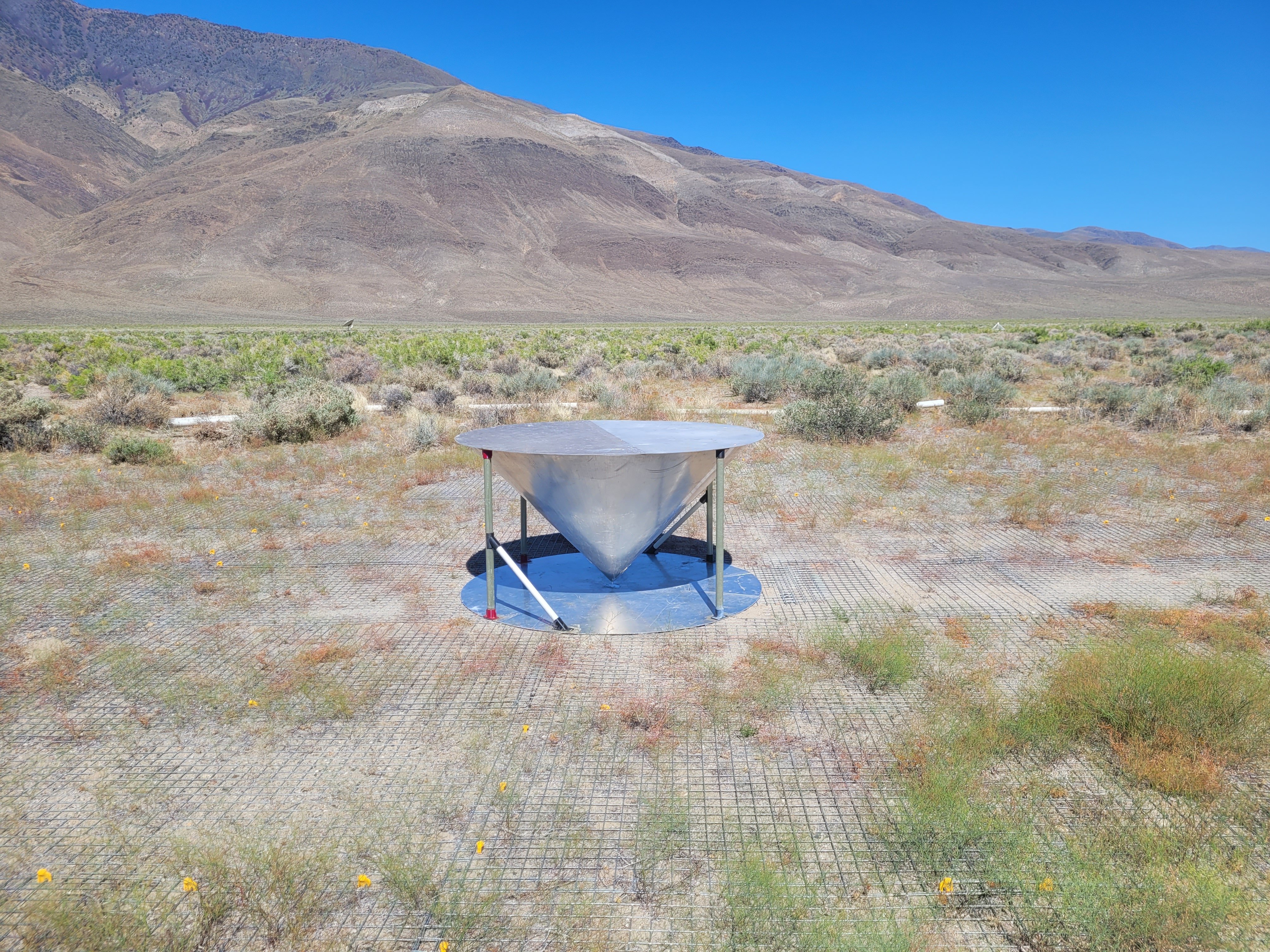}
    \caption{MANAS monopole comprising of the conical antenna and its circular ground plane. The antenna system sits on an extended serrated ground screen 20m by 20m (tip-to-tip) to improve its gain.}
    \label{fig:1}
\end{figure}

The signal chain begins at a disc-cone monopole antenna, whose achromatic profile is scaled from the SARAS design\cite{2021arXiv210401756N}. A monopole antenna was chosen for a number of reasons. It eliminates the need for a balun since its unbalanced terminals allow direct coaxial connection while the finite conductive ground plane provides natural horizon nulls for RFI mitigation from terrestrial sources. Careful design of the monopole yields spectrally smooth reflection and radiation characteristics with a smoothly varying beam, preserving the spectral smoothness of the foregrounds\cite{2021arXiv210401756N}. Electromagnetic simulations of the antenna (FEKO) show a single broad beam pointed at zenith whose width grows with frequency (Figure~\ref{fig:beamcuts}a), varying smoothly across the band without resonant features that would imprint structure on the recovered spectrum (Figure~\ref{fig:beamcuts}b). Figure~\ref{fig:1} shows the deployed antenna on its ground screen.

The antenna feeds an EDGES-2 receiver\cite{monsalve_calibration_2017} mounted at its base. During observing, the receiver switches between the antenna and internal reference loads to track and remove time-variable gain fluctuations in the signal chain; the absolute temperature scale is set separately by a prior laboratory calibration of the receiver against four known
loads. The receiver is powered over the same coaxial cable that carries the RF (RF+DC), which runs $\sim$350 m to the instrument shelter, where the signal is fed to the MANAS analog backend and digitizer.
In the planned configuration, a 3 dB splitter will route the signal along a second path to the OVRO-LWA backend. Reading the antenna out simultaneously through the absolutely calibrated MANAS receiver and the interferometer will allow its beam to be mapped in situ against the array, and the array's continuous ionospheric
monitoring to be applied to MANAS.

The digital backend is a Xilinx RFSoC ZCU111 board running CASPER firmware. It samples at 491.52 MS/s and applies a 65{,}536-point polyphase filterbank, yielding
32{,}768 spectral channels at 7.5 kHz resolution across the 0--245.76 MHz Nyquist band, with spectra transferred over a 10G link to a Supermicro 7049A-T workstation
for control, storage, and processing. Antenna and receiver reflection coefficients ($S_{11}$) are measured in situ with a Keysight FieldFox VNA driven through a
four-position switch under LabJack control.

\begin{figure}
    \centering
    \includegraphics[width=0.45\linewidth]{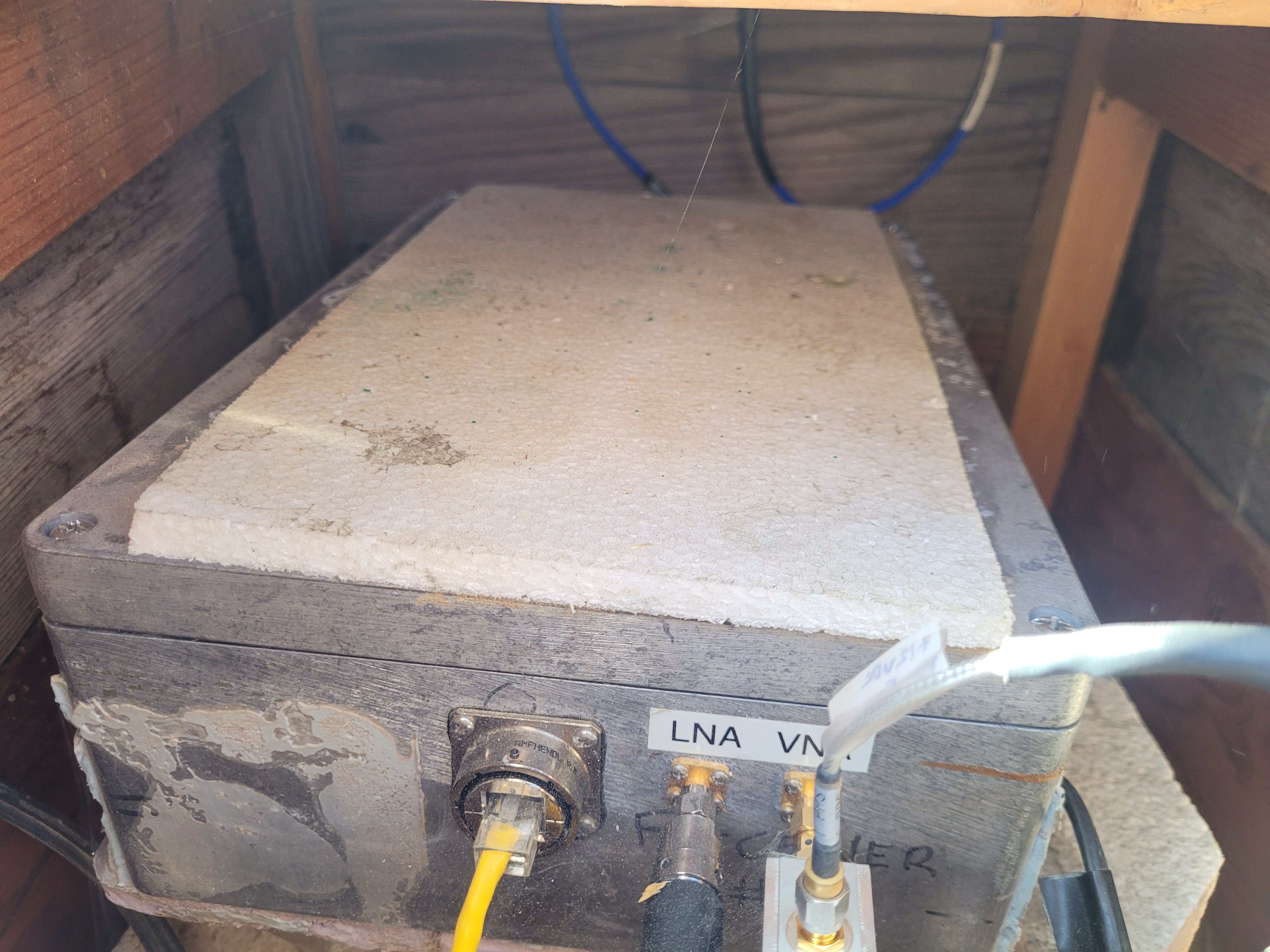}
     \includegraphics[width=0.45\linewidth]{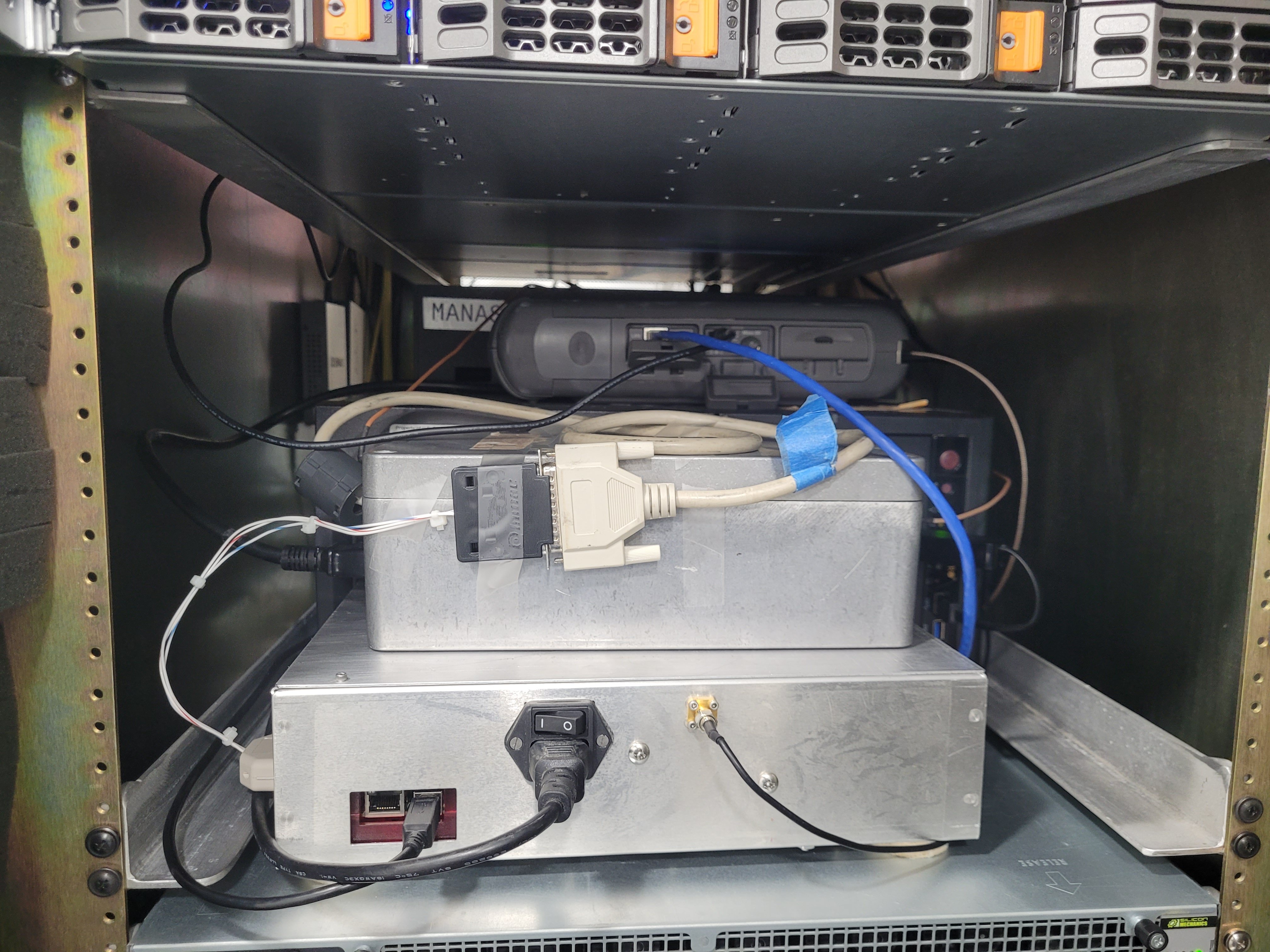}
    \caption{[left]The EDGES receiver under the ground plane. The receiver is temperature-controlled by two pipes. One has a fan that extracts hot air while the other one brings in cold air. [right] The backend, workstation, switching board, and VNA in their rack; located in the shelter 350m away from the antenna.}
    \label{fig:2}
\end{figure}

\begin{figure}
    \centering
    \includegraphics[width=0.9\linewidth]{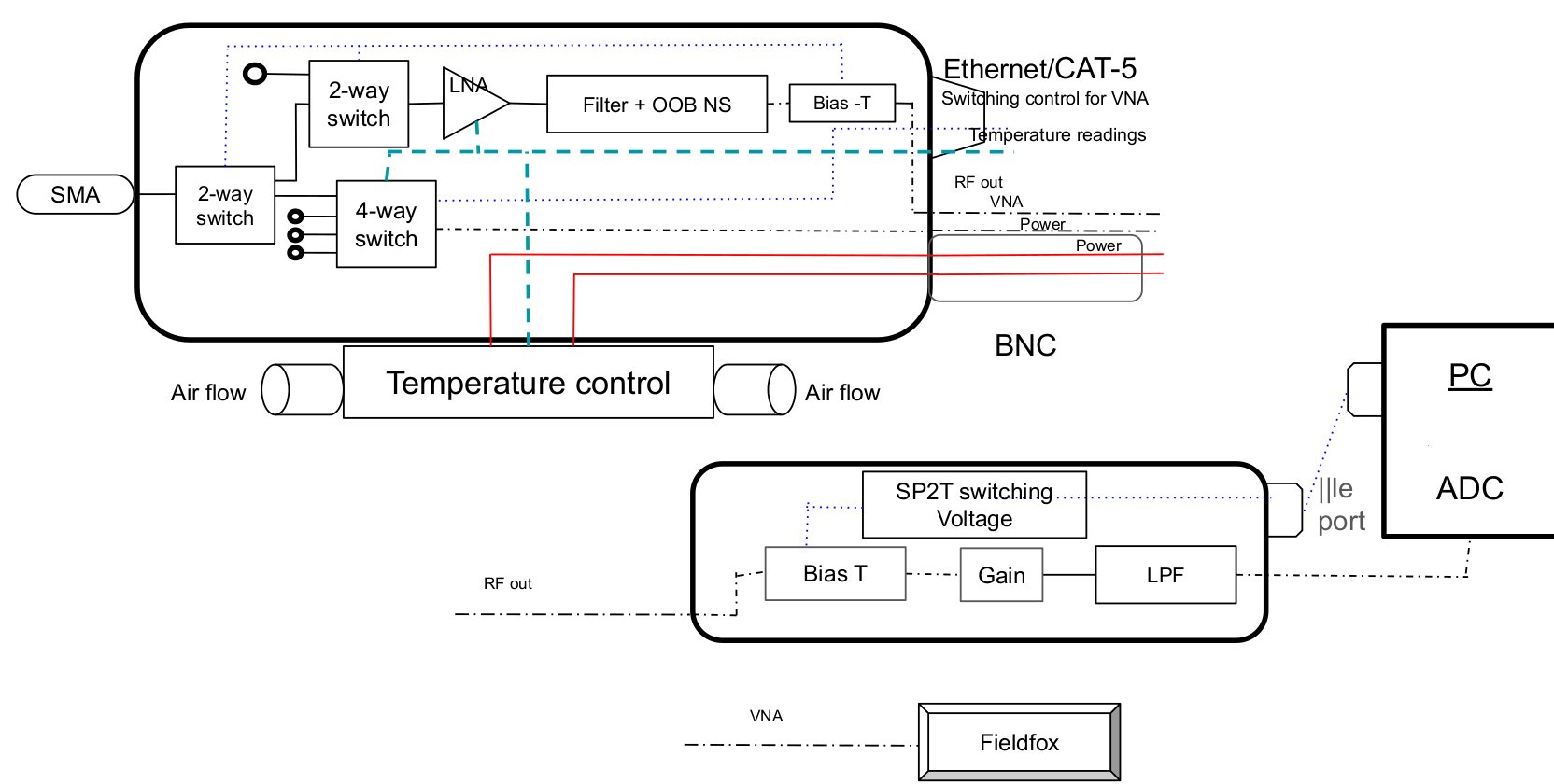}
    \caption{Block diagram of the EDGES receiver MANAS uses. It highlights the dicke switching for gain calibration and 4-way for the in-situ S11 measurements.}
    \label{fig:placeholder}
\end{figure}

\pagebreak
\section{CURRENT STATUS}
In October 2025, we replaced one of the OVRO-LWA outrigger antennas, element 252, with the MANAS monopole. This pad was chosen for three reasons: it lies $\sim$200~m from the array core, far enough to limit coupling to the dense central array; it is one of the five coaxial-fed outriggers, matching the MANAS signal chain; and it retains the large ground plane left over from the LEDA experiment. The antenna structure is supported on fiberglass stands (visible in Figure \ref{fig:1}); being non-conductive and effectively RF-transparent, they support the assembly without perturbing the antenna beam or introducing spurious reflections. The antenna port uses a custom-built Type~N connector with an air dielectric. Maintaining a 50~$\Omega$ impedance with an air dielectric, rather than a solid one, requires a larger body and a thicker center conductor, which is mechanically well-matched to the antenna terminal.

The receiver is mounted in a buried enclosure beneath the antenna, accessed through a hatch in the ground plane (Figure~\ref{fig:2} [left]). It is actively
temperature-controlled by two ducted pipes: one with a fan that extracts warm air, the other drawing in cool air which holds the receiver temperature stable. This
matters because both the receiver gain and the calibration solution vary with temperature. The fan sits in a separate below-ground enclosure located 10~m from the antenna at the edge of the extended ground plane, keeping it well away from the antenna phase center. It is powered by a DC voltage carried on the VNA cable, combined and separated using bias tees, so the same line carries both the $S_{11}$ signal and fan power. The receiver itself is powered over a separate coaxial cable that carries the RF (RF+DC). 

Two switching circuits were developed for MANAS. The first provides the three-position switching used for gain calibration: during spectrometer acquisition, the RF input is cycled between the antenna and internal references (noise source on and noise source off), so that time-variable gain in the signal chain can be tracked and removed. The second is a four-position switch that connects either the antenna or one of three reflection standards: open, short, and matched load to the VNA for the in-situ $S_{11}$ measurements. Both switches are driven by a LabJack T7 digital I/O module under computer control: the acquisition software commands the LabJack over its host link, and the LabJack's digital output lines set the switch states. The three-position calibration switching is fed into the EDGES analog backend and carried to the receiver on the RF cable, while the four-position $S_{11}$ control is sent on a twin lead cable across the field terminating on an CAT5 Ethernet pigtail that in turn manages the power distribution inside the receiver.

\begin{figure}[h!]
    \centering
    \includegraphics[width=0.6\linewidth]{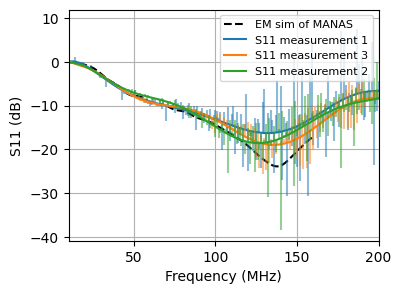}
    \caption{Measured antenna $S_{11}$ from three in-situ acquisitions (colored) compared with the EM simulation (dashed). The measurements track the simulation across the band; the increased scatter above $\sim$130~MHz reflects reduced return power over the $\sim$350~m cable for low-reflection terminations.}
    \label{fig:s11}
\end{figure}

The digital backend and server are housed in the OVRO-LWA shelter. A 10 MHz reference clocks the ADC on the RFSoC while the 1 PPS will be hooked up to synchronize the instrument with the OVRO-LWA array before the in situ beam mapping and calibration gets under way.

The instrument achieved first light in June 2026, with the antenna, receiver, and spectrometer integrated end-to-end and producing on-sky spectra. The end-to-end system evaluation is well under way with daily data dumps from the RFSoC spectrometer. First on-sky data confirm a strong radio-frequency interference (RFI) environment. Daytime spectra are dominated by the FM broadcast band; nighttime data, collected when the band is quieter, is being used to characterize the usable
RFI-free spectrum.

In-situ reflection coefficient ($S_{11}$) measurements of the antenna are taken periodically with the FieldFox VNA via the switching circuit. The VNA is accessible through the workstation which allows for measurements to be taken remotely. The measured $S_{11}$ agrees well with the electromagnetic simulations across the band (Figure~\ref{fig:s11}). Within the 30–88~MHz band of interest the $S_{11}$ varies smoothly, with the reflection minimum near 130~MHz falling outside the band, so no resonant feature is imprinted on the measurement over the science band. The agreement degrades at the upper frequencies, where the trace becomes noisy and develops a ripple - an effect traced to the $\sim$350~m cable run between the antenna and the shelter. 

Once continuous, autonomous operation and routine data-quality checks are in place, the first-light data will be calibrated with \texttt{pygsdata}\footnote{\url{https://github.com/edges-collab/pygsdata}} to estimate the Galactic
spectral index, providing an end-to-end check of the instrument and its absolute calibration. Beyond this, work will proceed toward the OVRO-LWA integration described above: splitting the antenna signal to the
array backend, mapping the MANAS beam in situ through holography on bright pulsars, incorporating the array's ionospheric calibration, and ultimately placing the
OVRO-LWA sky maps on an absolute flux scale.



\acknowledgments 
We thank the staff at the Owens Valley Radio Observatory for their assistance during the instrument's assembly and testing. We are grateful to Jonathon Kocz for
providing the ZCU111 board and the associated spectrometer acquisition software. This work was supported by Moffett funds donated to the California Institute of Technology.
\bibliography{report} 
\bibliographystyle{spiebib} 

\end{document}